\documentclass[useAMS,usenatbib,usegraphicx]{mn2e}
\usepackage{graphicx,times}

\title[A new disc tracer for \protect{B[e]} supergiants]{A new observational tracer for 
high-density disc-like structures around B[e] supergiants\thanks{Based on 
observations collected with the ESO 2.2m-telescope in La Silla, Chile, under
programme 076.D-0609(A)}}
\author[A. Aret et al.]{A.~Aret,$^{1,2}$\thanks{E-mail: 
aret@aai.ee (AA); kraus@sunstel.asu.cas.cz (MK); 
fmuratore@carina.fcaglp.unlp.edu.ar (MFM); borges@on.br (MBF)} 
M.~Kraus,$^{1}$\footnotemark[2] M.F.~Muratore$^{3,4}$\footnotemark[2] and 
M.~Borges Fernandes$^{5}$\footnotemark[2]\\
$^{1}$ Astronomick\'y \'ustav, Akademie v\v{e}d \v{C}esk\'e republiky,
Fri\v{c}ova 298, 251\,65 Ond\v{r}ejov, Czech Republic\\ 
$^{2}$ Tartu Observatory, 61602, T\~oravere, Tartumaa, Estonia\\
$^{3}$ Departamento de Espectroscop\'ia Estelar, Facultad de Ciencias
Astron\'omicas y Geof\'isicas, Universidad Nacional de La Plata,\\
~~~Paseo del Bosque s/n, B1900FWA, La Plata, Argentina\\
$^{4}$ Instituto de Astrof\'isica de La Plata, CCT La Plata, CONICET-UNLP,
Paseo del Bosque s/n, B1900FWA, La Plata, Argentina\\
$^{5}$ Observat\'orio Nacional, Rua General Jos\'e Cristino 77,
20921-400 S\~ao Cristov\~ao, Rio de Janeiro, Brazil
}

\begin{document}

\date{Accepted 2012 March 2. Received 2012 March 2; in original form 2011 December 13}

\pagerange{\pageref{firstpage}--\pageref{lastpage}} \pubyear{2012}

\maketitle

\label{firstpage}

\begin{abstract}
The disc formation mechanism of B[e] supergiants is one of the puzzling phenomena 
in massive star evolution. Rapid stellar rotation seems to play an important 
role for the non-spherically symmetric mass-loss leading to a high-density disc or 
ring-like structure of neutral material around these massive and luminous objects. 
The radial density and temperature structure as well as the kinematics within this
high-density material are, however, not well studied. Based on high-resolution 
optical spectra of a sample of B[e] supergiants in the Magellanic Clouds we 
especially searched for tracers of the kinematics within their discs. Besides the 
well-known [O\,{\sc i}] lines, we discovered the [Ca\,{\sc ii}] $\lambda\lambda$7291,
7324 lines that can be used as a complementary set of disc tracers. We find that
these lines originate from very high-density regions, located closer to the star than
the [O\,{\sc i}] $\lambda$5577 line-forming region. The line profiles of both
the [O\,{\sc i}] and the [Ca\,{\sc ii}] lines indicate that the discs or rings of 
high-density material are in Keplerian rotation. We estimate plausible ranges of disc 
inclination angles for the sample of B[e] supergiants and suggest that 
the star LHA\,120-S\,22 might have a spiral arm rather than a disc.
\end{abstract}


\section{Introduction}

B[e] supergiants (B[e]SGs) are evolved massive stars with large amounts of 
circumstellar material in the form of atomic and molecular gas as well as dust.
The presence of gas is indicated by the numerous emission lines from both
forbidden and permitted transitions of atoms in different ionization stages.
Molecular emission has been reported for many B[e]SGs: TiO emission has been
detected at optical wavelengths 
\citep[Torres et al., in preparation]{Zickgraf89} and CO emission has been observed 
in the near-infrared
\citep*{McGregor, Morris, Liermann, Muratore}. 
The presence of dust is obvious from the strong infrared excess emission in
the photometric and spectroscopic data \citep[e.g.,][]{Zickgraf86, Bonanos09, 
Bonanos10}, and the 
composition of this dust shows indication for both oxygen (crystalline
silicates) and carbon (polycyclic aromatic hydrocarbons, PAHs) chemistry
\citep{Kastner10}.

The co-existence of a normal B-supergiant radiation driven wind and a 
large amount of molecular and dusty material suggests that the geometry of the 
circumstellar material cannot be spherically symmetric, and polarimetric 
studies of B[e]SGs in the Magellanic Clouds \citep[e.g.,][]{Magalhaes, 
Magalhaesetal, Melgarejo} and of Galactic B[e] stars and B[e]SG candidates 
\citep{OudmaijerDrew}
supported a flat, disc-like geometry. Though studied for many decades, the 
structure and geometry, as well as the formation history 
of the B[e]SGs' discs still remain unknown.

\citet{Zickgraf85} suggested that the discs are formed by a slow, high-density
equatorial outflow. Such an equatorially outflowing disc could result from
rapid stellar rotation via the bi-stability mechanism \citep*{LamersPauldrach,
Pelupessy} or from the newly found slow-wind solutions \citep*{Cure04, Cure05}
that produce a slow but high-density wind in the equatorial region. Rapid
stellar rotation could also be the reason for recombination of the ionized
wind material in the equatorial plane \citep[e.g.,][]{Kraus06}, resulting in
large amounts of neutral hydrogen close to the stellar surface, as was
recently found for some B[e] stars and B[e]SGs \citep*{KrausBorges, Kraus07, 
Kraus10}.

\begin{table*}
\begin{minipage}{120mm}
\caption{Parameters of the observed B[e]SGs.}              % title of Table
\label{tab:parameters}      % is used to refer this table in the text
\begin{tabular}{l c c c c c c}
\hline                        % inserts one horizontal line
Object & Sp. Type & $E(B-V)$ & $T_{\rm eff}$ & $L$ & Orientation & References \\    % table heading
 &  & & (10$^3$K) & (10$^5L_{\odot}$) &  &  \\    % table heading
\hline                                   % inserts single horizontal line
\multicolumn{7}{c}{SMC} \\[3pt]
%\hline                                   % inserts single horizontal line
 LHA 115-S 18  & B0      & 0.40 & 25    & 3-4.6 & $\pm$pole-on & (1) \\ % inserting body of the table
 LHA 115-S 65  & B2-3    & 0.15-0.20 & 17    & 5.0   & edge-on	& (2) \\[3pt]
%\hline
\multicolumn{7}{c}{LMC} \\[5pt]
%\hline
 LHA 120-S 12  & B0.5    & 0.20-0.25 & 23    & 2.2   & intermediate & (2) \\
 LHA 120-S 22  & B0-B0.5 & 0.25-0.30 & 23-26 & 6-9.5 & edge-on	& (2) \\
 LHA 120-S 73  & B8      & 0.12 & 12    & 2.9   & $\pm$pole-on & (2) \\
 LHA 120-S 111 &         & 0.28 & 15.5  & 11.2  &		& (4) \\
 LHA 120-S 127 & B0.5    & 0.25 & 22.5  & 13    & pole-on	& (3) \\
 LHA 120-S 134 & B0      & 0.20-0.25 & 26    & 7.9   & pole-on	& (2) \\

\hline      
\end{tabular}

\medskip
References: (1) \citet{Zickgraf89}; (2) \citet{Zickgraf86}; (3) \citet{Zickgraf85}; (4) \citet{McGregor}.\\

\end{minipage}
\end{table*}

However, for some individual B[e]SGs recent detailed investigations 
revealed that the simple outflowing-disc scenario does not hold.
The set of [O{\sc i}] $\lambda \lambda$5577, 6300 lines trace
physically distinct regions within the inner parts of the massive disc or ring
which are so dense that hydrogen is already predominantely neutral
\citep{Kraus07, Kraus10}. 
On the other hand,
information on the density, temperature, and kinematics within the molecular
parts of the disc can be obtained from the strength and structure of the CO
first-overtone bands \citep[e.g.,][]{Carr93, Chandler, Carr95, Najita, Kraus00,
Liermann, Muratore}.
The fluxes, temperatures, and kinematics obtained from these emissions
hint towards the presence of a detached ring (or detached disc-like structure)
rather than a disc structure formed from a steady outflow \citep{Liermann,
Muratore, Kraus10}. 

To study the formation mechanisms of B[e]SGs' discs it is important to combine
the information of density, temperature and especially kinematics obtained
from different sources, like optical emission lines and infrared molecular bands that
trace different regions within the circumstellar material. The forbidden emission
lines play a very important role in this study, because they are optically thin 
\citep[e.g.,][]{Osterbrock} and hence ideal tracers for density and temperature
\citep[see e.g.,][]{Kraus05}, and their profiles contain the full kinematical 
information of their formation region \citep{Ignace, Kraus10}. However, most 
forbidden emission lines are formed in the low-density
regions. The chance to find forbidden lines originating from the
high-density disc complementing the well-known set of the [O{\sc i}] lines 
is rather low. Also, high-quality optical and near-infrared observations for 
most of the B[e]SGs are still lacking.

To overcome the difficulties in studying the structure and formation history
of the B[e]SGs' discs we started an observational campaign focused on the 
search for typical density and velocity tracers in the spectra of these stars.
Here we report on our results based on high-resolution optical spectra 
obtained for a total of eight B[e]SGs in the Magellanic Clouds. In these 
spectra we discovered besides the well-known [O\,{\sc i}] lines a set of 
forbidden emission lines, the [Ca\,{\sc ii}] $\lambda\lambda$7291, 7324 lines,
as a new, valuable disc tracer.

\section{Observation and reduction}
\label{sec:obs}
Our sample consists of two B[e]SGs from the Small Magellanic Cloud and six from 
the Large Magellanic Cloud (SMC and LMC, respectively). The objects are listed in 
Table\,\ref{tab:parameters} together with the stellar parameters and possible disc
orientations as found in the literature. 

The observations have been carried out on two consecutive nights, on 2005 
December 10 and 11. We obtained high-resolution optical spectra using the Fiber-fed
Extended Range Optical Spectrograph (FEROS) attached to the 2.2-m telescope
at the European Southern Observatory in La Silla (Chile). 

FEROS is a bench-mounted Echelle spectrograph with two fibres, each of them
covering a circular sky area of 2$\arcsec$ in diameter. The wavelength coverage
extends from 3600\,\AA \ to 9200\,\AA \ and the spectral resolution is $R =
55\,000$ (in the region around 6000\,\AA). We made our own data reduction, since 
the pipeline presented some problems during that mission. Based on this, we used 
available \textsc{midas} routines \citep{Verschueren}, including some modifications related 
to the FEROS characteristics. Thus, our data are reduced differentially relative 
to a master stellar and flat-field frame, considering the usual steps for echelle 
data reduction with the proper location and merge of the spectral orders. The 
spectra are rebinned taking into account the natural resolution of the detector. 
The telluric and barycentric velocity corrections have also been performed.
From the sky lines we measured the accuracy of the
wavelength calibration, which is better than 0.5\,km\,s$^{-1}$.

To identify and delete cosmic rays from the spectra, we took a minimum of two
spectra per star. The spectral lines did not show any variability in the individual 
exposures, so we added them up to achieve a better signal-to-noise ratio (S/N).
Exposure times together with the final achieved signal-to-noise ratios are listed in
Table 2. The stellar radial velocities obtained from the central 
wavelength of the [O\,{\sc i}] $\lambda$6300 line are also given in that table.

\begin{table}
 \caption{Exposure times (number of exposures is given in parentheses) and signal-to-noise ratios at 5500 {\AA}. The radial velocities $V_r$ were derived using the [O\,{\sc I}] $\lambda$\,6300.}
      \label{tab:snr}
\begin{center}
  \begin{tabular}{@{\extracolsep{10pt}}@{}lrcc@{}}
  \hline
Object         & Exposure (s) & Total S/N  & $V_r$ (km\,s$^{-1}$)\\
\hline
LHA 115-S 18  &1800 (2)      &50           & 146 	\\
              & 900 (1)      &             &     	\\
LHA 115-S 65  & 900 (2)      &55           & 189 	\\
LHA 120-S 12  & 900 (2)      &60           & 290	\\
LHA 120-S 22  & 900 (2)      &70           & 289	\\
LHA 120-S 73  & 600 (1)      &100          & 261	\\
              & 900 (1)      &             &    	\\
LHA 120-S 111 & 900 (2)      &80           & 267	\\
LHA 120-S 127 & 450 (2)      &50           & 259	\\
LHA 120-S 134 &1800 (1)      &70           & 267	\\
              & 900 (1)      &             &   	\\
\hline
\end{tabular}
\end{center}
\end{table}

\section{Results}
\label{sec:results}

The FEROS spectra of the B[e]SGs display a large number of emission lines from
permitted and forbidden transitions. Emission lines from the latter are
especially valuable, because forbidden lines are optically thin, and their
profiles reflect the kinematics within their formation region. One interesting
set of lines has been identified as the forbidden emission lines of neutral
oxygen, [O\,{\sc i}] $\lambda\lambda$5577, 6300, 6364. While the two lines
$\lambda\lambda$6300, 6364 arise from the same upper energy level and hence are
formed in the same region within the disc, the [O\,{\sc i}] $\lambda$5577 line
emerges from regions of higher density \citep[for details see][]{Kraus07,
Kraus10}, i.e., from distances very close to the star.
The ratio [O\,{\sc i}] $\lambda$6300/$\lambda$5577 serves as an ideal
indicator for temperature, density, and ionization fraction of the disc
material \citep{Kraus07, Kraus10}. Since the emission of these two individual
lines arises from physically distinct disc regions, i.e., regions with different
velocities, their line fluxes and profiles deliver complementary information.
This makes these two lines a
valuable tool in the study of both the structure and kinematics of the discs,
as was recently shown by \citet{Kraus10} for the B[e]SG LHA\,115-S\,65 in the
SMC.

\begin{figure*}
\includegraphics[width=0.95\textwidth]{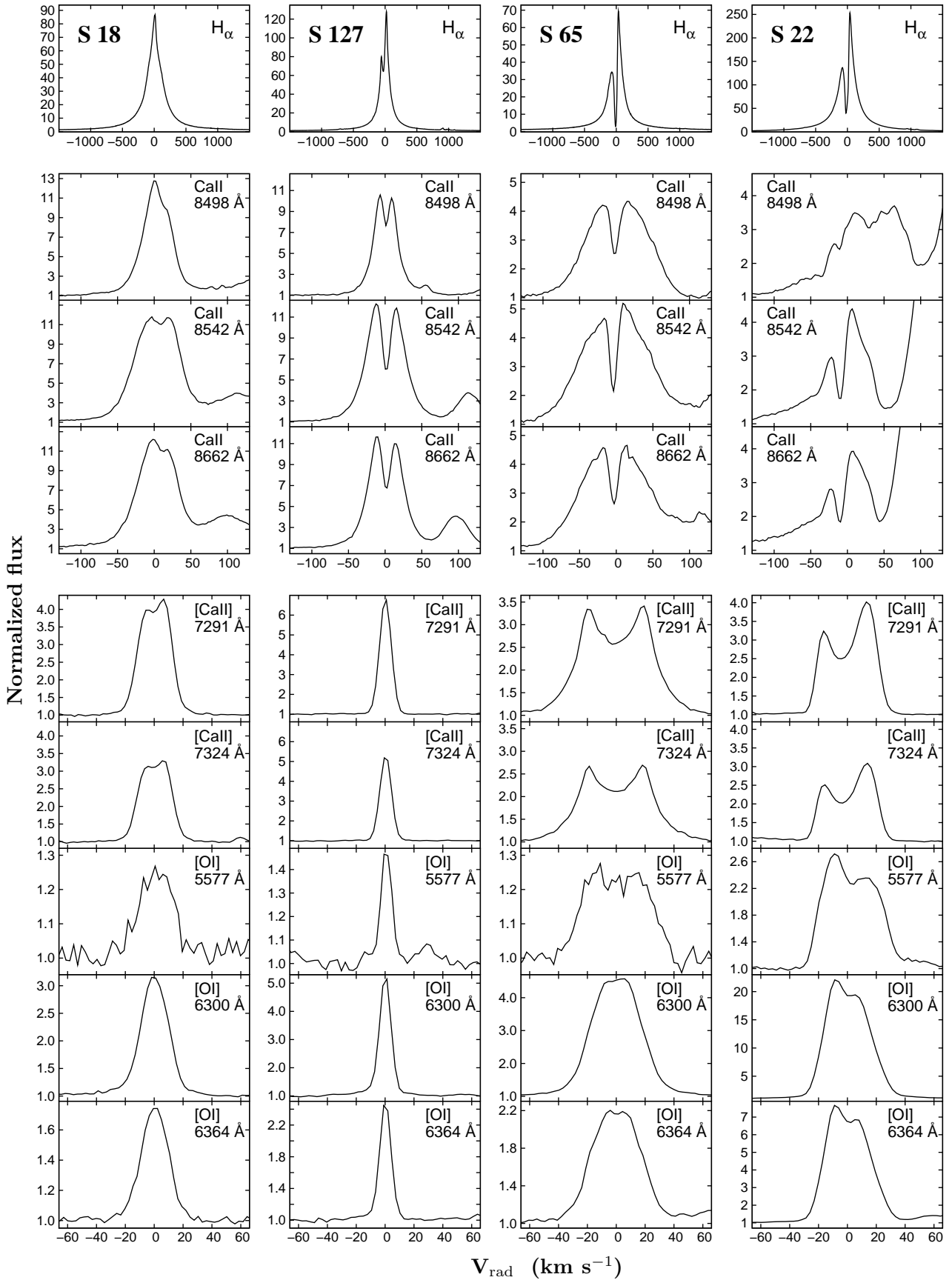}
\caption{Line profile variety of stars seen pole-on (S\,18, S\,127) and edge-on
(S\,65, S\,22). For details see text.}
\label{pole-edge}
\end{figure*}

\begin{figure*}
\includegraphics[width=0.95\textwidth]{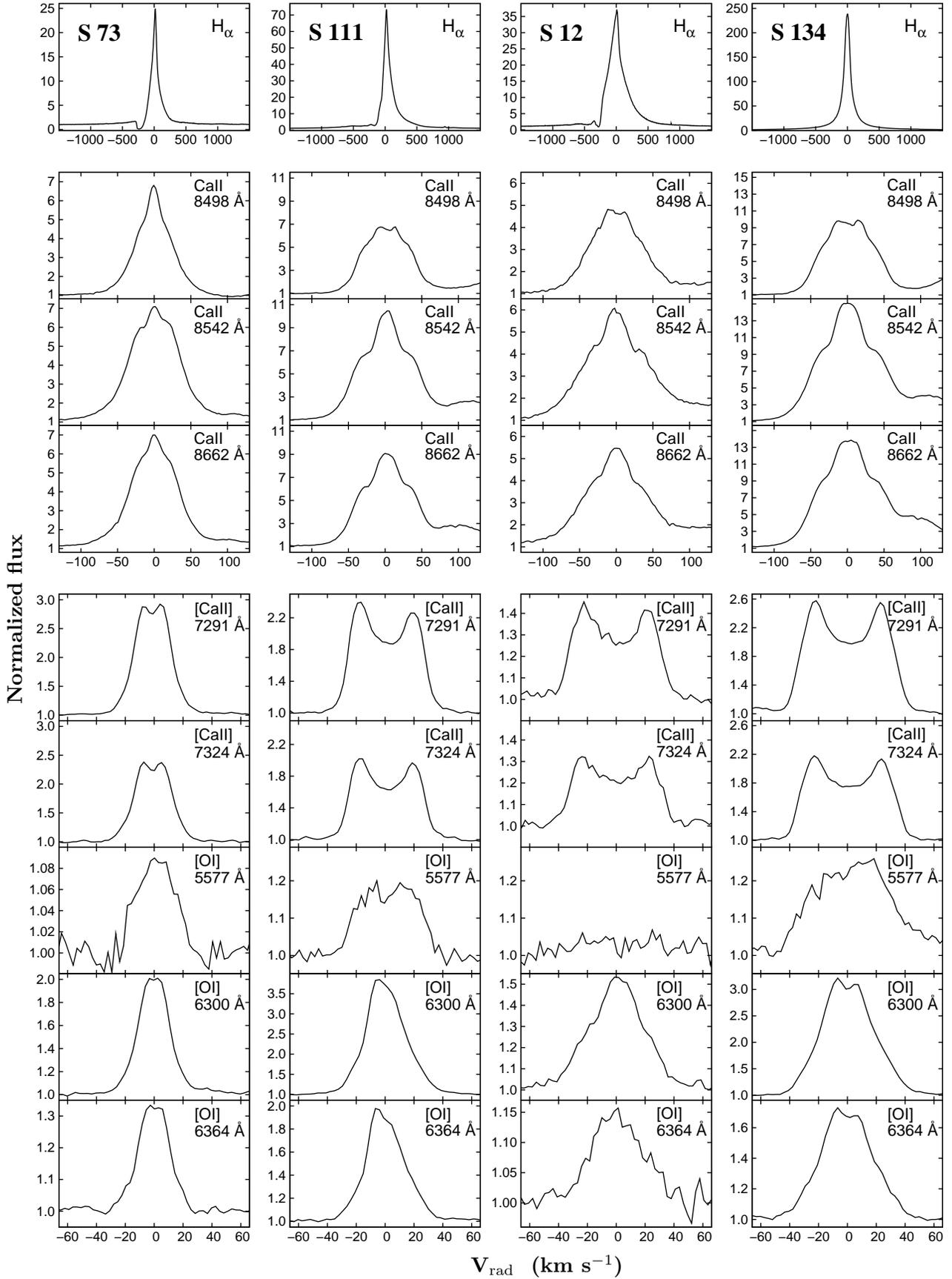}
\caption{Same as Fig.\,\ref{pole-edge} but for stars with intermediate orientation.}
\label{intermediate}
\end{figure*}

Unfortunately, for many stars of our sample the [O\,{\sc i}] $\lambda$5577 line
is quite weak or even absent (see Figs\,\ref{pole-edge} and 
\ref{intermediate}). In addition, the S/N ratio of the observed spectra in the
wavelength region of this line is rather low (see Table\,\ref{tab:snr}). 
This hampers severely a proper determination of the 
physical parameters of their discs and emphasizes the need for additional
lines that are capable of tracing the disc properties at different distances from
the star.

\begin{figure*}
\includegraphics[width=0.95\textwidth]{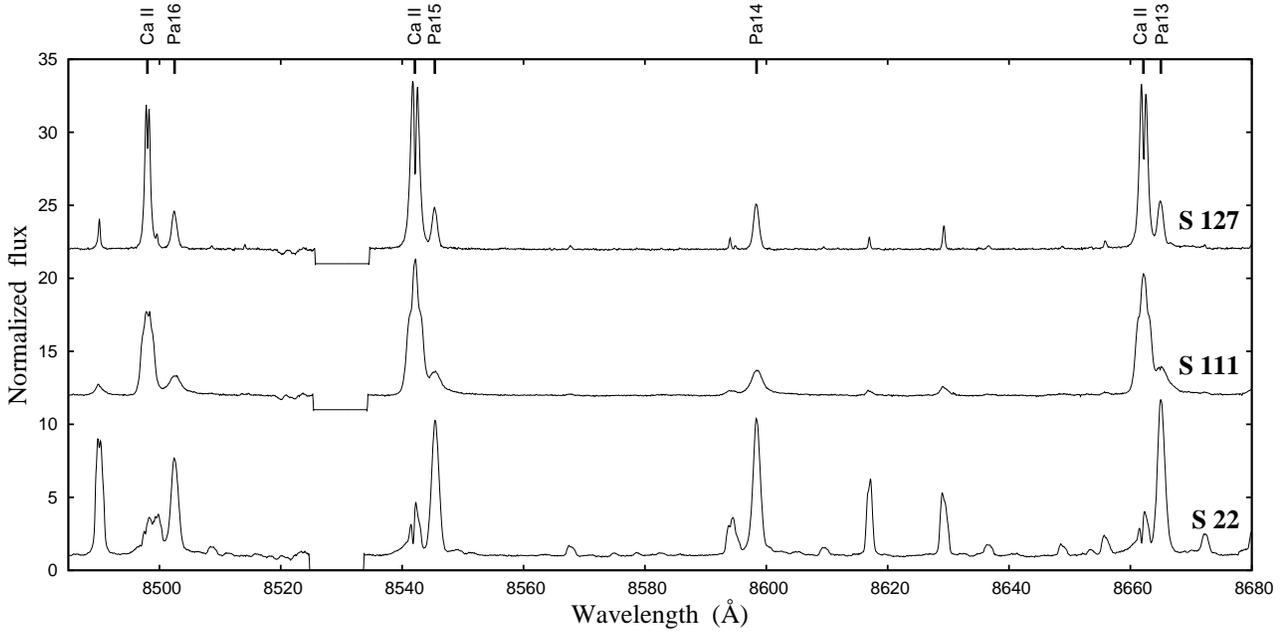}
\caption{Ca\,{\sc ii} infrared triplet in LHA\,120-S\,22, LHA\,120-S\,111 
and LHA\,120-S\,127. The spectra of LHA\,120-S\,111 and LHA\,120-S\,127 
were shifted upwards for a better viewing (by 11 and 21, respectively). The gap around 
8530\,\AA~is a short wavelength region that is not covered by the red orders of FEROS.} 
\label{triplet}
\end{figure*}

Upon inspection of the FEROS spectra we found that all our B[e]SGs display
strong line emission from the Ca\,{\sc ii} $\lambda\lambda$8498, 8542, 8662
infrared triplet. Examples of the FEROS spectra around these lines are 
shown in Fig.\,\ref{triplet} for three stars of our sample with different 
Ca\,{\sc ii} line profiles and different strengths in the adjacent hydrogen lines of the Paschen series.
This emission is known to arise from high-density regions, and 
circumstellar discs are especially favoured locations. Emission in the Ca\,{\sc ii} 
infrared triplet has been reported for a diversity of objects: T\,Tauri stars
\citep[e.g.,][]{HamannPerssonI, Kwan}, Herbig Ae/Be stars \citep[e.g.,][]
{HamannPerssonII}, classical Be stars \citep*[e.g.,][]{Polidan, Briot, Jaschek,
Andrillat88, Andrillat90}, and other, non-supergiant B[e] stars
\citep{Marcelo01, Marcelo09}.

Furthermore, we discovered strong line emission in [Ca\,{\sc ii}] 
$\lambda\lambda$7291, 7324 in all our objects. So far, the appearance of
these forbidden emission lines has been reported for only a few other disc
sources \citep*{Hamann, Hartigan}. Nevertheless, the detection of these
forbidden lines is of particular interest, because the energy levels from which
the [Ca II] lines emerge are coupled to those from which the Ca II infrared
triplet lines arise. The presence or absence of the [Ca II] lines could thus be
a suitable tracer for specific temperature and density conditions within the 
disc or wind.

In Figs\,\ref{pole-edge} and \ref{intermediate} we plot the line profiles
of the calcium lines together with those of the [O\,{\sc i}] lines
and of H$\alpha$. To compare widths and shapes of lines from different stars we
plotted identical lines on the same radial velocity scale.
% H$\alpha$ from
%$-1500$\,km\,s$^{-1}$ to $+1500$\,km\,s$^{-1}$, the lines of the Ca\,{\sc ii}
%infrared triplet from $-130$\,km\,s$^{-1}$ to $+130$\,km\,s$^{-1}$, and the
%forbidden lines ([O\,{\sc i}], [Ca\,{\sc ii}]) from $-66$\,km\,s$^{-1}$ to
%$+66$\,km\,s$^{-1}$. 
Stellar radial velocities as listed in
Table\,\ref{tab:snr} have been subtracted. The close vicinity of the 
Ca\,{\sc ii} infrared triplet lines to adjacent Paschen lines could result 
in line-blending. The distances of the individual
Paschen lines Pa(16) $\lambda$8502, Pa(15) $\lambda$8545 and Pa(13) $\lambda$8665,
from the Ca\,{\sc ii} lines are $+157.7$\,km\,s$^{-1}$, 
$+115.5$\,km\,s$^{-1}$ and $+99.7$\,km\,s$^{-1}$, respectively. 
However, in the high-resolution FEROS spectra the shape of the Ca\,{\sc ii} lines 
is not distorted by the Paschen lines for most of our stars (see 
Figs\,\ref{pole-edge}, \ref{intermediate}, and \ref{triplet}).
One exception is LHA\,120-S\,22, in which the Paschen lines are extremely strong.

\begin{table*}
\vskip -10pt
\begin{minipage}{0.94\textwidth}
\caption{Shapes and peak separations (km s$^{-1}$) of forbidden calcium and
oxygen lines.}
\label{tab:forbidden}
\begin{tabular}{@{\extracolsep{10pt}}@{}l*{10}{c}@{}}
\hline
Object & \multicolumn{2}{c}{[Ca\,{\sc ii}] 7291 \AA}&
\multicolumn{2}{c}{[Ca\,{\sc ii}] 7324 \AA}& \multicolumn{2}{c}{[O\,{\sc i}]
5577 \AA}& \multicolumn{2}{c}{[O\,{\sc i}] 6300 \AA}&
\multicolumn{2}{c}{[O\,{\sc i}] 6364 \AA}\\
& Shape & Sep. & Shape & Sep. & Shape & Sep. & Shape & Sep. & Shape & Sep.\\
\hline
LHA 115-S 18 & double & 10 & double & 10 & single & -- & single & -- &
single & --\\
LHA 115-S 65 & double & 38 & double & 37 & double? & -- & double & n.\,r. &
double & 8\\
LHA 120-S 12 & double & 42 & double & 45 & no line & -- & single & -- &
single? & --\\
LHA 120-S 22 & double & 30 & double & 30 & double & 20 & double & 13 &
double & 14\\
LHA 120-S 73 & double & 11 & double & 12 & single & -- & double & 6 & double
& 6\\
LHA 120-S 111 & double & 36 & double & 36 & double? & -- & double & n.\,r. &
double & n.\,r.\\
LHA 120-S 127 & single & -- & single & -- & single & -- & single & -- &
single & --\\
LHA 120-S 134 & double & 45 & double & 46 & broad single? & -- & double & 13
& double & 13\\
\hline
\end{tabular}

n.\,r. -- not resolved.
\end{minipage}
\end{table*}

\subsection{Line profile diversity}

The appearance of the optical spectra of the B[e]SGs has been described in 
detail in the literature \citep[e.g.,][]{Zickgraf85, Zickgraf86, Zickgraf89}. 
However, former observations were limited to wavelengths shorter than 
7000\,{\AA}. This is the reason why the strong emission in both the Ca\,{\sc ii} infrared triplet 
and the forbidden [Ca\,{\sc ii}] lines was not reported earlier. Hence, we refrain
from repeating a detailed spectral description, and instead we focus on these
newly detected emission lines and discuss 
their shape with respect to the [O\,{\sc i}] lines.
% and H$\alpha$.

The different line widths of the forbidden lines are expected to reflect the 
kinematics of physically distinct regions. We note an increase in 
line width starting with the [O\,{\sc i}] $\lambda\lambda$6300, 6364 lines, 
through the [O\,{\sc i}] $\lambda$5577 line, to the [Ca\,{\sc ii}] $\lambda\lambda$7291, 
7324 lines for all stars but LHA\,120-S\,127. In addition, we notice a 
simultaneous change in line profile. If the [O\,{\sc i}] $\lambda\lambda$6300, 
6364 lines have a single-peaked, Gaussian shape, the [Ca\,{\sc ii}] $\lambda
\lambda$7291, 7324 lines show slight indications for double-peaks. If the 
[O\,{\sc i}] $\lambda\lambda$6300, 6364 lines show slightly double-peaked 
profiles, the [Ca\,{\sc ii}] $\lambda\lambda$7291, 7324 lines are wide and clearly 
doubly-peaked. The width and shape of the [O\,{\sc i}] $\lambda$5577 line is 
intermediate. However, a clear classification of its profile shape is difficult
for most of our objects, because this line is usually weak and noisy. 
The profile shapes and the measured (if possible) peak-separations in 
velocities are listed in Table\,\ref{tab:forbidden}.

The lines of the Ca\,{\sc ii} infrared triplet are always (much) broader than the
forbidden emission lines, and their line profiles show a diversity ranging from
double (Fig.\,\ref{pole-edge}) to multiple-peaked (Fig.\,\ref{intermediate}).
Where possible, the peak-separation in velocity has been measured and listed in
Table\,\ref{tab:IR}. The double-peaked profiles of these lines, however, do
not resemble those structures that refer to the formation within a rotating or
equatorially outflowing medium as it seems to be the case for the forbidden
lines. Instead, they probably stem from absorption within a high-density
medium.

The emission in the H$\alpha$ line is the most prominent feature in all our spectra. 
Its strength ranges from about 25 to about 250 times the continuum flux. Its profile 
can be single-peaked, double-peaked, or of P\,Cygni type. Its wings extend for all
objects to large velocities. These high velocities, however, do not reflect
real kinematics but are most probably caused by electron scattering in a
high-density wind or disc.

\subsection{Disc structure and kinematics}

The obvious diversity we see in the line profiles indicates that the formation
regions for different lines must be physically \mbox{distinct}. This allows us to
draw some (though preliminary) conclusions about the possible disc kinematics 
and density structure of the sample stars.

As was recently shown by \citet{Kraus07,Kraus10}, the [O\,{\sc i}] $\lambda$5577 
line arises from regions of much higher density and hence closer to the star than 
the [O\,{\sc i}] $\lambda\lambda$6300, 6364 lines. Thus, besides information
on temperature and density this set of lines provides details on the kinematics 
within the two 
line-forming regions. For the B[e]SG star LHA\,115-S\,65 which is viewed edge-on
the [O\,{\sc i}] $\lambda$5577 line is broader than the other two [O\,{\sc i}] 
lines and with stronger indication of a double-peaked profile. Simultaneous
modelling of the [O\,{\sc i}] line profiles and line luminosities led to the 
conclusion that the disc around LHA\,115-S\,65 is detached from the star and is in 
Keplerian rotation \citep{Kraus10}. Furthermore, this detached disc or ring has 
a rather small outflow velocity which decreases with distance from the star. 
These findings are in strict contrast to the general belief that B[e]SG stars' 
discs are the high-density equatorial outflow parts within the two-component 
wind scenario of \citet{Zickgraf85}.

Comparison of the [O\,{\sc i}] line profiles for all our sample stars indicate
that the $\lambda$5577 line is indeed almost always the broadest one suggesting
that Keplerian rotation rather than outflow might hold for all B[e]SG discs.
Exceptions are the stars LHA\,120-S\,127 that is viewed exactly pole-on so that 
all [O\,{\sc i}] lines are equally broad and LHA\,120-S\,12
for which the [O\,{\sc i}] $\lambda$5577 line could not be detected. If the discs
around the B[e]SGs are indeed in Keplerian rotation, the broader and clearly
double-peaked line profiles of the [Ca\,{\sc ii}] lines must 
originate from regions closer to the star than the [O\,{\sc i}] $\lambda$5577 
line-forming region. Consequently, the density in the [Ca\,{\sc ii}] line-forming
region must be higher.

\begin{table}
\caption{Shapes and peak separations (km s$^{-1}$) of Ca\,{\sc ii} infrared
triplet.}
\label{tab:IR}
\begin{tabular}{@{}l*{6}{c}@{}}
\hline
Object & \multicolumn{2}{c}{Ca\,{\sc ii} 8498 \AA}&
\multicolumn{2}{c}{Ca\,{\sc ii} 8542 \AA}& \multicolumn{2}{c}{Ca\,{\sc ii}
8662 \AA}\\
& Shape & Sep. & Shape & Sep. & Shape & Sep. \\
\hline
LHA 115-S 18 & double & n.\,r. & double & 23 & double & 19 \\
LHA 115-S 65 & double & 33 & double & 26 & double & 31 \\
LHA 120-S 12 & multi & -- & triple & -- & triple & -- \\
LHA 120-S 22 & multi & -- & double & 28 & double & 28 \\
LHA 120-S 73 & triple & -- & triple & -- & triple & -- \\
LHA 120-S 111 & multi & -- & triple & -- & triple & -- \\
LHA 120-S 127 & double & 16 & double & 17 & double & 25 \\
LHA 120-S 134 & multi & -- & triple & -- & triple & -- \\
\hline
\end{tabular}

n.\,r. -- not resolved.
\end{table}

Recently, \citet{Hartigan} investigated the [Ca\,{\sc ii}] $\lambda$7291/7324
line ratios in both the low- and the high-density limit and found that the
ratio changes only a little from 1.495 in the former to 1.535 in the latter.
They suggested the critical electron density to be on the order of $\sim
5\times 10^{7}$\,cm$^{-3}$. To obtain the [Ca\,{\sc ii}] $\lambda$7291/7324
flux ratio, we measured the equivalent widths of the lines. 
We supposed that the real flux ratio does not differ substantially from the
measured ratio of the  equivalent widths because the lines have a separation
of only 30\,\,{\AA}. To confirm this, we checked the influence of the different
continuum strengths and the effect of dereddening due to interstellar extinction 
by applying Kurucz model atmospheres and using the extinction values 
from the literature as given in Table\,\ref{tab:parameters}. 
We found that for all our stars corrections to the measured equivalent width ratios 
are less than 1\%. The final values for the [Ca\,{\sc ii}]
$\lambda$7291/7324 line ratios range from 1.33 to 1.47.
All these values fall below the value
of 1.495 for the low-density limit found by \citet{Hartigan}.
Thus, we cannot determine the electron density in
the [Ca\,{\sc ii}] line forming region from the observed flux ratios.
Instead, we may use the values found within the
[O\,{\sc i}] $\lambda$5577 line-forming region as lower limits for the electron
densities within the [Ca\,{\sc ii}] line-forming regions. However, so far proper
density determinations within the [O\,{\sc i}] line-forming regions have been
performed for only one object, namely LHA\,115-S\,65 \citep{Kraus10}.

The SMC star LHA\,115-S\,65 was reported to be viewed edge-on \citep{Zickgraf86,
Kraus10}. Its [Ca\,{\sc ii}] lines are much broader than the [O\,{\sc i}] lines and
clearly double-peaked. The electron density within the [O\,{\sc i}] $\lambda$5577
line-forming region was found to be $n_{\rm e} \simeq 9\times 10^{6}$\,cm$^{-3}$
\citep{Kraus10}.
The range in rotational velocities over which the [Ca\,{\sc ii}] lines seem to be
formed spreads from about 52\,km\,s$^{-1}$ as measured from the extent of the
line wings to about 19\,km\,s$^{-1}$ as obtained from the peak-separation of
their profiles. Current studies of the evolutionary link of B[e]SGs to other
phases in massive star evolution indicate that they could be evolving back 
bluewards, which means that they are in a post-red supergiant (post-RSG) or 
post-yellow hypergiant (post-YHG) phase \citep[][Muratore et 
al. in preparation]{Muratore}. Applying Keplerian rotation for the stellar mass 
obtained for such a scenario and given in Table\,\ref{tab:incl}, 
we find that the velocity of 52\,km\,s$^{-1}$ corresponds to a distance of
about 22\,$R_{*}$ and the velocity of 19\,km\,s$^{-1}$ to about 160\,$R_{*}$.
Thus, most of the emission originates from the disc regions at roughly 160\,$R_{*}$.
If we assume that the disc has formed from stellar 
mass-loss, then its radial density structure has a $r^{-2}$ distribution 
according to the law of mass conservation. Consequently,
the density increases by about a factor of $\ga 50$ over the [Ca\,{\sc ii}]
emission region (from the outer to the inner edge) and becomes (much) higher
than the critical density, meaning that we are in the high-density regime of
the [Ca\,{\sc ii}] line formation.

The rather large range in density might be one reason why the values of the
[Ca\,{\sc ii}] flux ratios do not agree with either of the limiting values computed
by \citet{Hartigan}. However, it should be noted that their values have been
determined under the assumption that the material is ionized and free electrons are
the dominant collision partners. Such a scenario does not hold for the discs of the
B[e]SGs. Instead, the hydrogen ionization fractions were found to be very small
within the [O\,{\sc i}] line-forming regions (but still delivering an electron
density higher than the critical density for [Ca\,{\sc ii}]). The discs around
B[e]SGs can thus be considered as being predominantely neutral in hydrogen
\citep{Kraus07, Kraus10}. Hence, collisions with hydrogen atoms certainly dominate
over collisions with free electrons and need to be considered together with a proper
radiation transfer in the lines of the infrared triplet. A proper modelling of the
Ca\,{\sc ii} atom is beyond the scope of the current work. Nevertheless, we may
claim that the [Ca\,{\sc ii}] lines provide a new
important and appropriate tracer for the structure and kinematics of B[e]SG stars'
discs.

\begin{figure*}
\includegraphics{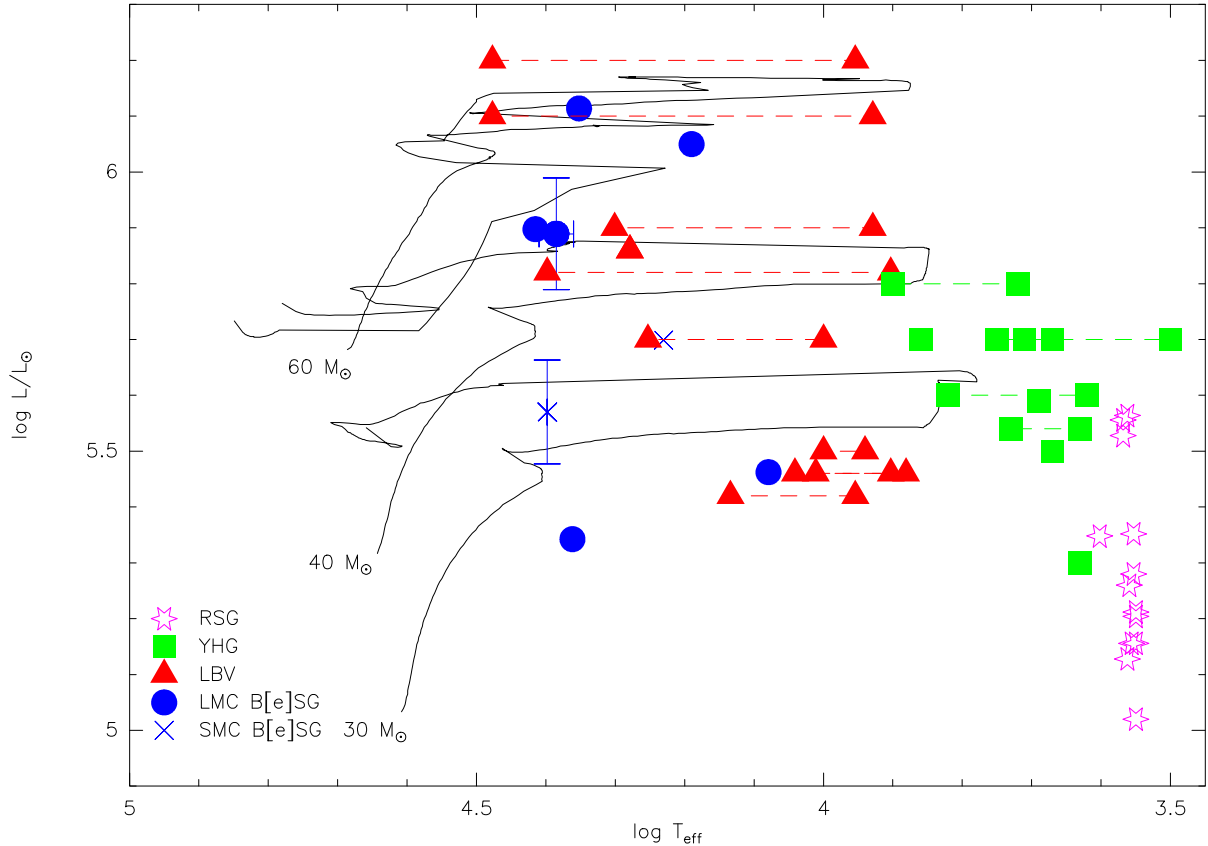}
\caption{Our B[e]SG sample in comparison to other massive post-main sequence phases. 
Stellar parameters are taken from \citet{HD} and \citet{Groh} for the LBVs, from 
\citet{deJager} for the YHGs, and from \citet{Levesque} for the RSGs. 
Stellar evolutionary tracks for rotating stars at LMC metallicity from 
\citet{MeynetMaeder05} are overplotted.}
\label{fig:evol}
\end{figure*}

\subsection{Masses of B[e]SGs -- their evolutionary phase}

The mass of LHA\,115-S\,65 used for the determination of the distance of and the density
within the [Ca\,{\sc ii}] line-forming region was obtained considering that the star is
evolving bluewards in the Hertzsprung-Russell diagram (HRD). In such a scenario, 
it could have lost the mass that is now 
forming its disc during the passage through a RSG and/or YHG phase (see Fig.\,\ref{fig:evol}).
So far, not much is known about the real evolutionary phase of B[e]SGs and several suggestions
about their present stage can be found in the literature. For instance, it has been suggested
that B[e]SGs are the results of massive binary mergers \citep*{Langer, Podsiadlowski}.
As of now such a scenario seems to hold for only one object, the SMC stars R4.
However, one should avoid to associate all B[e]SGs with stellar mergers, because the
spectral characteristics of R4, its position in the HRD just at the end of
the main sequence, and its strong photometric variability, which is not common for B[e]SGs,
displaces R4 from the rest of the B[e]SG sample. We therefore consider the merger scenario as
the least likely common origin for the B[e] phenomenon in the supergiant stars.

Some B[e]SGs were suggested to be evolutionary linked to Luminous Blue Variables (LBVs) 
with which they share about the same region in the HRD (see Fig.\,\ref{fig:evol}). 
For instance, the star V39 in the low metallicity galaxy IC\,1613 
was classified by \citet{Herrero} as either a B[e]SG or an LBV candidate, 
and the B[e]SG star LHA\,115-S\,65 was classified by \citet{Kraus10} as a pre-LBV.
The LBVs are luminous objects ($\log L/L_{\odot} \ge 5.4$). They
were found to separate into two groups, and this separation happens at
$\log L/L_{\odot} \approx 5.8$. While the more luminous LBVs were suggested to
be post-main sequence stars evolving redwards, their lower luminosity
counterparts are thought to evolve bluewards, i.e., they are suggested to be
post-RSGs or post-YHGs \citep[][see also Fig.\,\ref{fig:evol}]{HD, Meynet11}.

Half of our objects certainly fall into this low-luminosity regime, while
two others are just at the border, but could still fit to the same
scenario because their location in the HRD indicates that they might have
evolved from a 40-42\,$M_{\odot}$ progenitor star, which is about the limiting
mass between the two evolutionary scenarios for LBVs described by
\citet{Meynet11}.
The possible classification  of at least one of our sample B[e]SGs as a pre-LBV
further suggests that the bluewards evolution in this low-luminosity regime
could indeed go from the YHG over the B[e]SG to the LBV phase.

In recent investigations \citet{Kastner10} studied the dusty discs of the
Magellanic Cloud B[e]SGs. They found convincing analogies to the discs around lower
mass stars in a post-asymptotic giant branch phase, so they concluded that the
B[e]SGs are most likely the more massive counterparts in a post-RSG (or post-YHG) phase.
And further support for such a scenario comes from recent investigations
of the $^{13}$C footprint for the two least massive B[e]SGs in our sample
(LHA\,120-S\,73 and LHA\,120-S\,12) revealing that a post-YHG scenario is most
likely for LHA\,120-S\,73 \citep{Muratore}. The $^{12}$C/$^{13}$C ratio found for 
LHA\,120-S\,12 was in agreement only with an evolved nature of the star. Whether 
LHA\,120-S\,12 could also be in a post-YHG phase needs to be
tested based on predictions of the $^{12}$C/$^{13}$C ratio from evolutionary
models for rotating stars at LMC metallicity extending to lower
initial masses than provided by the current models of \citet{MeynetMaeder05},
but such models are not yet available.

However, a post-YHG scenario certainly fails for the most luminous B[e]SGs, for which no
YHG phase exists. At which stage in the evolution they might be (just leaving the main-sequence
as was suggested for the most luminous LBVs or on a blue loop) is not known.

\begin{table*}
\begin{minipage}{170mm}
\caption{Stellar masses, Keplerian rotation velocities, and inclination angles as estimated from the [Ca\,{\sc ii}] line profiles. Values of the disc mass fluxes, $f_{d}$, are taken from \citet{habil}.} 
\label{tab:incl}
\begin{tabular}{@{}l*{9}{c}@{}}
\hline
Object & $M_{\rm ZAMS}$ & $M_{\rm current}$ & $R_{*}$ & $f_d/f_{d({\rm S65})}$ & $r$([Ca\,{\sc ii}]) &
  $v_{\rm Kep}(r)$ & $v_{\rm los}$ & $r(v_{\rm Kep} = v_{\rm los})$ 
  & suggested $i$\\
& ($M_{\odot}$) & ($M_{\odot}$) & ($R_{\odot}$) &  & ($R_{*}$) & (km\,s$^{-1}$) &
  (km\,s$^{-1}$) & ($R_{*}$) & ($^{\circ}$)\\
\hline
LHA 115-S 18  & $29\pm 2$ & $19\pm 3$ & $33\pm 3$  & 2.15 & 235 & $21.7\pm 2.7$ & 5 & $4491\pm 1104$ & $13.5\pm 2$ \\ 
LHA 115-S 65  & $35\pm 1$ & $25\pm 1$ & 82         & 1.00 & 160 & $19.0\pm 0.4$ & 19 & $161\pm 7$ & $84\pm 6$ \\
LHA 120-S 12  & $22\pm 2$ & $14\pm 2$ & 30         & 2.15 & 235 & $19.8\pm 1.0$ & 21-22.5 & $191\pm 41$ & $\sim 90$ \\
LHA 120-S 22  & $41\pm 4$ & $25\pm 4$ & $49\pm 11$ & 3.85 & 314 & $18.0\pm 3.5$ & 15 & $469\pm 173$ & $67\pm 23$ \\
  & $49\pm 4^{\rm a}$ & $41\pm 4$ &  &  &  & $23.1\pm 3.7$ &  & $763\pm 240$ & $42\pm 8$ \\
LHA 120-S 73  & $25\pm 1$ & $15\pm 1$ & 125        & 0.78 & 141 & $12.7\pm 0.4$ & 6 & $634\pm 42$ & $28\pm 1$ \\
LHA 120-S 111 & $60\pm 5$ & $40\pm 5$ & 147        & unknown & -- & -- & 18 & $171\pm 12$ &  \\
 &  $60\pm 5^{\rm a}$ & $48\pm 5$ &        &  & &  &  & $192\pm 20$ &  \\
LHA 120-S 127 & $\ge 60$ & unknown & 75            & 1.31 & 183 & -- & unresolved & -- & $\sim$0 \\
LHA 120-S 134 & $42\pm 1$ & $26\pm 1$ & 44         & 2.46 & 251 & $21.2\pm 0.5$  & 23 & $196\pm 8$ & $\sim 90$ \\
 & $50\pm 2^{\rm a}$ & $42\pm 2$ &          &  & & $26.9\pm 0.6$  &  & $344\pm 16$ & $59\pm 2$ \\
\hline
\end{tabular}

$^{\rm a}$ assuming the star has just left the main-sequence.
\end{minipage}
\end{table*}

\subsection{Disc inclination angles}

We now return to the [Ca\,{\sc ii}] lines. The measured peak separation of their line profiles 
(see Table\,\ref{tab:forbidden}) can be used to estimate possible ranges in disc inclination angles.
For this, we need to know the appropriate distance from the stars at which the bulk of 
the [Ca\,{\sc ii}] lines form. For
LHA\,115-S\,65 we found that this distance is about $160\,R_{*}$. The density reached
at that distance should be very similar for the [Ca\,{\sc ii}] line-forming region
for all B[e]SGs. We can estimate these distances based on the disc mass fluxes
that were determined by \citet{habil} 
%from the $J$ excess 
for all our sample stars
but LHA\,120-S\,111. Both the ratios of disc mass fluxes of the sample stars with respect
to LHA\,115-S\,65 and the resulting distances at which the required density 
%for the bulk of [Ca\,{\sc ii}] line-forming region 
is reached are given in Table\,\ref{tab:incl}.

With the obtained distances, inclination angles can be estimated from the measured
peak-separations together with the assumption of a Keplerian rotation law. To
calculate Keplerian rotation velocities the current stellar
masses need to be known. In Fig.\,\ref{fig:evol} we show the positions of our objects
in the HRD together with evolutionary tracks from
\citet{MeynetMaeder05} computed for rotating stars at LMC metallicity. The evolutionary
tracks for SMC metallicity do not differ substantially from those for the LMC, so we
refrain from plotting both sets of tracks. As discussed in the previous section, 
the B[e]SGs with $\log L/L_{\odot} \la 5.8$ might be in a post-YHG phase, while
for the most luminous ones it is not clear whether they are evolving to the red or 
to the blue side of the HRD.

Assuming that B[e]SGs are evolving bluewards and interpolating available evolutionary tracks, 
we obtained zero-age main-sequence and current masses for all objects as listed in 
Table\,\ref{tab:incl}. For the most massive stars in the sample we also list current 
masses obtained under the assumption that
these stars might just have evolved off the main-sequence. The stellar
radii have been obtained from the temperature and luminosity values given in
Table\,\ref{tab:parameters}.

The line-of-sight velocities, $v_{\rm los}$, result from the peak-separations of the 
[Ca\,{\sc ii}] lines. If all stars were seen edge-on, these velocities would
correspond to the Keplerian velocities. To get an idea about maximum distances of
the [Ca\,{\sc ii}] line-forming regions from the central stars we calculated 
the radial distances resulting from the peak-separation velocities.
These values are listed in Table\,\ref{tab:incl} in terms of stellar radii.
Obviously, these distances are quite large for those stars that are considered to
be seen more or less pole-on like, e.g., LHA\,115-S\,18.

We computed the Keplerian rotation velocities for all objects at the distance 
$r$([Ca\,{\sc ii}])
where the bulk of the [Ca\,{\sc ii}] line emission is thought to originate. 
From the ratio of the observed rotational velocity projected to the 
line-of-sight and the expected rotational velocity, we obtained a range in inclination 
angles. These values are included in Table\,\ref{tab:incl}.
Good agreement with formerly
suggested orientations is achieved for most objects. In particular, the stars
LHA\,115-S\,18 and LHA\,120-S\,73 are found to be $\pm$pole-on. 

The inclination angle of LHA\,120-S\,111 has not yet been determined. The position
of this star in the HRD shows that it belongs to the most luminous objects. The masses estimated 
for both evolutionary scenarios can only be considered as rough estimates, and the missing value
of its disc mass flux hampers a proper inclination determination. Nevertheless, considering
the mass ranges as listed in Table\,\ref{tab:incl}, maximum distances of the [Ca\,{\sc ii}] 
line-forming region as obtained for a possible edge-on orientation are only slightly larger
than the one obtained for LHA\,115-S\,65. The line strengths of the [O\,{\sc i}] and 
[Ca\,{\sc ii}] lines are comparable to those of the other B[e]SGs, so that its disc mass flux
should not be considerably smaller compared to LHA\,115-S\,65. Hence, we might conclude that
LHA\,120-S\,111 could be oriented close to edge-on.

The values obtained for the B[e]SGs at the border between low- and high-luminosity
regime, LHA\,120-S\,22 and LHA\,120-S\,134, differ significantly
for the two different evolutionary scenarios; while a close to edge-on orientation is
obtained for the post-YHG scenario, an intermediate inclination angle is obtained for 
the redwards evolutionary scenario.

For LHA\,120-S\,134, both results clearly disagree with the pole-on orientation
(see Table\,\ref{tab:parameters}) suggested by \citet{Zickgraf86}. The broad and 
clearly double-peaked line profiles of its [Ca\,{\sc ii}] lines definitely speak against 
a pole-on orientation (see Fig.\,\ref{intermediate}). 
Which scenario will turn out to be the correct one needs further detailed investigation.

The orientation of LHA\,120-S\,22 in the redwards evolutionary scenario 
delivered a value of only $i \simeq 42\pm 8\degr$. But this object has been reported 
to be viewed edge-on \citep{Zickgraf86}. This orientation has been confirmed by
\citet{Chu} who discovered a reflection nebula in the vicinity of LHA\,120-S\,22
that mirrors different viewing angles. Interestingly, a (close-to) edge-on 
orientation is found in the post-YHG scenario. 

Inspection of the line profiles of LHA\,120-S\,22 reveals that this object has 
another peculiarity: both sets of forbidden lines are double-peaked and clearly
asymmetric. But while the [Ca\,{\sc ii}] lines show a much stronger red peak, the
[O\,{\sc i}] lines display a stronger blue peak. Considering that the 
[Ca\,{\sc ii}] lines are supposed to originate from distances closer to the star 
than the [O\,{\sc i}] lines, this behaviour indicates 
the presence of inhomogeneities in the circumstellar material like 
a density wave in the disc or a spiral arm-like 
structure. Such structures have recently been seen on images of the disc of a young 
stellar object\footnote{http://www.nasa.gov/topics/universe/features/possible-planets.html}. 
So, LHA\,120-S\,22 might be the first B[e]SG star with spiral arms. 
In any case, it is worth to study both the kinematics of its circumstellar material and 
the time variability of the line profiles in more detail.

We would like to emphasize that the inclination angles listed in
Table\,\ref{tab:incl} can only be considered as rough estimates, because
several assumptions have been made for their derivation. The one with the
highest uncertainty is the distance at which the [Ca\,{\sc ii}] line emission 
forms, because we implicitly adopt that the ionization fraction
found for LHA\,115-S\,65 is the same in all B[e]SG star discs. This must not be true. 
A higher value for the ionization fraction implies a higher electron density, and 
the distance at which the [Ca\,{\sc ii}] emission forms is shifted farther out. 
The Keplerian rotation velocities are lower there and consequently the inclination of the disc would be larger. Proper values for the ionization fractions in the discs of all B[e]SGs are therefore needed. This emphasizes the need of flux calibrated data for all the sample stars.

\section{Conclusions}

We report on the discovery of the [Ca\,{\sc ii}] $\lambda\lambda$7291, 7324 lines 
in high-resolution spectra of a sample of B[e] supergiants. These lines originate
from different regions than the already well-known disc tracers, the [O\,{\sc i}] lines.
The [Ca\,{\sc ii}] lines trace regions of higher density and thus must form  
closer to the star. 
Combination of the kinematical information obtained from the line 
profiles of the [O\,{\sc i}] and [Ca\,{\sc ii}] allows to suggest that the material
around the B[e] supergiants is in Keplerian rotation. We estimated disc inclinations 
which are in reasonably good agreement with former guesses.   
However, the clearly asymmetric line profiles seen in the spectra of the star 
LHA\,120-S\,22 indicate that the material is not in a disc or ring-like structure.
We suggest this star might have a spiral arm seen edge-on.

In addition, we discuss plausible evolutionary phases for the B[e]SGs.
Although perhaps for some individual B[e]SGs (like R4 in the SMC) a different
scenario might hold, a blueward evolution of (most of) the B[e]SGs seems
to be likely. As suggested by \citet{Kraus09} and proven by \citet{Liermann}
and \citet{Muratore}, the $^{13}$C footprint is a powerful tool to determine
the evolutionary phase of
B[e]SGs. Hence reliable data of the $^{13}$C footprint for all B[e]SGs are needed,
and we recently started an observational campaign to study the $^{13}$C enrichment
in the discs of all B[e]SGs. Results from these observations will certainly
provide a better basis for a more detailed discussion about the evolutionary phase of
B[e]SGs and their possible links to other evolved phases of massive star evolution.

\section*{Acknowledgments}

A.A. and M.K. acknowledge financial support from GA\,\v{C}R under grant numbers 
205/08/0003 and 209/11/1198. This work was supported by the research project SF0060030s08 
of the Estonian Ministry of Education and Research. M.F.M. acknowledges Universidad Nacional de La Plata for the research fellowship.
M.B.F. acknowledges Conselho Nacional de Desenvolvimento Cient\'ifico e
Tecnol\'ogico (CNPq-Brazil) for the post-doctoral grant. M.B.F. acknowledges Herman Hensberge for all support with FEROS data reduction.

\bsp
\label{lastpage}
\end{document}